\documentclass[aps,floats,twocolumn,prd,showpacs,nofootinbib]{revtex4}

\usepackage{amssymb}
\usepackage{amsmath}

\usepackage{graphicx}
\usepackage{graphics}
\usepackage{dcolumn}
\usepackage{color}
\usepackage{rotate}
\usepackage{fancyhdr}
\usepackage{subfigure}

\def\be{\begin{equation}}
\def\ee{\end{equation}}
\def\bea{\begin{eqnarray}}
\def\eea{\end{eqnarray}}
\def\ba{\begin{eqnarray}}
\def\ea{\end{eqnarray}}

\definecolor{darkred}{rgb}{.743,0,0}

\begin{document}

\title{Cosmological perturbations without the
Boltzmann hierarchy}

\author{Marc Kamionkowski}
\affiliation{Department of Physics and Astronomy, Johns
     Hopkins University, 3400 N.\ Charles St., Baltimore, MD 21218}

\date{\today}

\begin{abstract}
Calculations of the evolution of cosmological perturbations
generally involve solution of a large number of coupled
differential equations to describe the evolution of the
multipole moments of the distribution of photon intensities and
polarization.  However, this ``Boltzmann hierarchy'' communicates
with the rest of the system of equations for the other
perturbation variables only through the
photon-intensity quadrupole moment.  Here I develop an
alternative formulation wherein this
photon-intensity quadrupole is obtained via solution of two
coupled integral equations---one for the intensity quadrupole
and another for the linear-polarization quadrupole---rather than
the full Boltzmann hierarchy. This alternative method of
calculation provides some physical insight and a cross-check for
the traditional approach.  I describe a simple and efficient
iterative numerical solution that converges fairly quickly.
I surmise that this may allow current state-of-the-art
cosmological-perturbation codes to be accelerated.
\end{abstract}

\maketitle

\section{Introduction}

Linear-theory calculations of the evolution of primordial
density perturbations provide the foundation
for the interpretation of cosmic microwave background and
large-scale-structure measurements.  They are thus an
essential tool in the construction of our current cosmological
model and in the continuing quest for new cosmological physics.

The calculations, which trace back over 50 years
\cite{earlywork}, involve time evolution of a set of coupled
differential equations \cite{1980s} for the metric perturbations
and for the dark-matter, baryon, neutrino, and photon density
and velocity perturbations.  There is also a (nominally
infinite) ``Boltzmann hierarchy'' of differential equations for
the higher moments (quadrupole, octupole, etc.) of the
photon-intensity and photon-polarization and
neutrino-momentum distributions.   The photon hierarchies can be
truncated at some maximum multipole moment $l_{\rm max}\simeq
30$ to provide sufficient precision for the monopole, dipole,
and octupole from which the higher-order moments (which provide the CMB
temperature/polarization power spectra) can be obtained through
a line-of-sight integral \cite{Seljak:1996is}.  Higher-order
extensions to the tight-coupling approximation (TCA)
\cite{CyrRacine:2010bk,Blas:2011rf}, improved numerical
integrators, and novel approximations to free-streaming
relativistic particles \cite{Blas:2011rf}) have provided
incredible code acceleration to what is still a fairly
complicated numerical calculation.  At present, virtually
all work in cosmology now relies on two publicly available codes,
{\tt CAMB} \cite{Lewis:1999bs} and {\tt CLASS}
\cite{Blas:2011rf}, which combine speed and precision with
model flexibility.

These codes are now extremely efficient and reliable.  However,
modern cosmological analyses, which employ Markov chain Monte
Carlos to map the likelihood in a multi-dimensional
parameter space, require these codes to be run repeatedly, thus
employing signficant computational resources.  It is thus
worthwhile to explore new numerical approaches.  New approaches
can also often provide new insights into the physics and may
perhaps provide tools that can be applied to other problems.

It was realized that for primordial tensor
perturbations (i.e., gravitational waves), the Boltzmann
hierarchy can be replaced by a small set of integral equations (IEs)
\cite{Weinberg:2003ur,Baskaran:2006qs}, an approach used in
Refs.~\cite{Flauger:2007es,Pritchard:2004qp}  A similar
approach was discussed for scalar perturbations (primordial density
perturbations) in Ref.~\cite{Weinberg:2006hh}, but not
implemented numerically.

Here, I re-visit this integral-equation approach for primordial 
density perturbations.  I discuss simplifications to the
equations in Ref.~\cite{Weinberg:2006hh} and describe a specific
implementation where the Boltzmann hierarchy for all photon
intensity/polarization multipole moments from the quadrupole
($l=2$) and higher are replaced by two IEs, one
for the photon quadrupole, and another for the polarization
quadrupole.  I discuss the numerical solution of these integral
equations and how the initial conditions for the IEs
are set from an early-time solution obtained with the TCA.  I
describe an iterative algorithm to solve these
integral equations simultaneously with the differential
equations for the other perturbation variables.  I show results
from two simple numerical codes that are identical except for the
replacement of the Boltzmann hierarchy in the first with the two
integral equations in the second.  Numerical experiments with
these codes suggest that this iterative IE algorithm may, with
further work, allow current state-of-the-art codes to be
accelerated.

This paper is organized as follows.  Section
\ref{sec:formalism}, presents and discusses the integral
equations.  Section \ref{sec:implementation} provides the
differential equations for the other perturbation variables
(i.e., for neutrinos, dark matter, baryons, and the metric) and
describe how the two integral equations are combined with these
other equations.  Section \ref{sec:numerics} describes a simple
algorithm to solve the integral equations numerically
and how the initial conditions for the IE solver are obtained
from the tight-coupling approximation at early times.
This Section also describes an iterative algorithm to solve them
in tandem with the differential equations.
Section~\ref{sec:results} describes the two rudimentary codes to
evolve the Boltzmann hierarchy and the IE equations.  I then
present and discuss results of the calculation.  Section
\ref{sec:conc} concludes with a discussion of possible concerns and
ideas for further steps.  Appendix \ref{sec:boltz} provides the photon
Boltzmann equations in the notation used here, and Appendix
\ref{sec:details} provides details of the algorithm to solve
the integral equation.  The codes are provided at {\tt
https://github.com/marckamion/IE} for readers interested to
follow up on calculational details that cannot be inferred from
the presentation here.

\section{Formalism}
\label{sec:formalism}

If we have a spectrum of initial curvature fluctuations with
power spectrum $P_{\cal R}(k) = \left\langle |{\cal R}_{\vec k}|^2
\right\rangle$, then the CMB temperature/polarization  power
spectra are
\begin{equation}
     C_l^{\rm XX'}= (2\pi^2)^{-1} \int\, k^2\, dk\,
     P_{\cal R}(k) \Delta_{kl}^{\rm X}(\tau_0) \Delta_{kl}^{\rm
     X'}(\tau_0),
\end{equation}
for X,X'$=$T,E with ``T'' the temperature and ``E'' the E-mode
of the polarization.  The transfer functions $\Delta_{kl}^{\rm
X}(\tau)$ are obtained through solution of
differential equations for the time evolution of the
relativistic gravitational potentials, the baryon,
dark-matter, photon, and neutrino densities and bulk velocities,
and the higher moments of the photon and neutrino momentum
distributions.  The moments of the intensity distribution of
photon momenta are the transfer functions $\Delta^{\rm
T}_{kl}(\tau)$ and the moments of the distribution of photon
polarizations are $\Delta^{\rm E}_{kl}(\tau)$.

The temperature transfer functions can be written as\footnote{The notation
here resembles largely that in
Ref.~\protect\cite{Blas:2011rf}.  The differences are that (i)
the photon $\Delta^{\rm T}_{kl}$ here is one quarter of theirs; (ii) the $R$
here is the inverse of their $R$; (iii) the $\dot\kappa$ here is their
$\tau_C^{-1}$; (iv) the $\alpha$ here is their $h+6\eta$.  The
$\Pi$ here is the same as that in Ref.~\cite{Seljak:1996is} and is
$\Pi=(F_{\gamma2}+G_{\gamma0}+G_{\gamma2})/4$ in terms of the
variables in Ref.~\cite{Blas:2011rf}.}
\begin{eqnarray}
     \Delta^{\rm T}_{kl}(\tau) &=&  \int_{\tau_i}^{\tau} \, d\tau' \,
     g(\tau,\tau') 
\left\{  \left[ -\frac16 \frac{\dot
     h_{k}(\tau')}{\dot \kappa(\tau')} + \Delta^{\rm T}_{k0}(\tau') \right]
     j_l( x) \right. \nonumber \\
     &- & \left.  \left[\frac13
     \frac{\dot\alpha_{k}(\tau')}{\dot\kappa(\tau')} + \frac12
      \Pi_{k}(\tau') \right] R^{\rm
     LL}_l(x) + \theta_{bk}(\tau') \frac{j_l'(x)}{k}
     \right\},\nonumber \\
\label{eqn:Tfluctuation} 
\end{eqnarray}
where $x=k(\tau-\tau')$; a dot denotes a partial
derivative with respect to $\tau$; and $g(\tau,\tau')= (d/d\tau')
e^{-\kappa(\tau,\tau')} = 
\dot\kappa(\tau')\, e^{-\kappa(\tau, \tau')}$ is the visibility 
function.  The initial conformal time $\tau_i$ must be taken to
be deep in the tight-coupling regime and will be discussed more
below.
Here, $\dot\kappa(\tau)=d\kappa/d\tau$ is the opacity,
the derivative of the Thomson-scattering optical depth with
respect to conformal time, and
\begin{equation}
     \kappa(\tau,\tau') = \int_{\tau'}^\tau \, d\tau_1\, \dot\kappa(\tau_1).
\end{equation}
Also, $R^{\rm LL}_l(x) = -\frac12 \left[ j_l(x)+ 3j_l''(x)
\right]$ \cite{Hu:1997hp,Dai:2012bc} in terms of spherical
Bessel functions $j_l(x)$, and
$\theta_{bk}(\tau)$ is the baryon velocity.  It is related
to the photon velocity (suppressing hereafter the subscript $k$
for notational economy) $\theta_\gamma(\tau)=3 k
\Delta^{\rm T}_{k1}(\tau)$ through
\begin{equation}
     \dot\theta_b = -{\cal H} \theta_b +c_s^2 k^2 \delta_b +
     \frac{\dot \kappa}{R} (\theta_\gamma-\theta_b),
\label{eqn:thetab}     
\end{equation}
where ${\cal H}(\tau)\equiv\dot a/a$ and $R(\tau)
\equiv (3/4) \rho_b(\tau)/\rho_\gamma(\tau)$, the scale factor
in units of $3/4$ of that at matter-baryon equality
($\rho_b(\tau)$ and $\rho_\gamma(\tau)$ are mean baryon and
photon energy densities, respectively).  The baryon sound speed $c_s$
is increasingly important on small scales but has little effect
on the larger distance/angular scales relevant for CMB fluctuations.
Here, $h(\tau)$ is the standard synchronous-gauge perturbation
variable, and $\alpha(\tau)=h(\tau)+6\eta(\tau)$ in terms of
the commonly used synchronous-gauge variable $\eta(\tau)$.

The function $\Pi(\tau)$ is a linear combination of the
photon-intensity and polarization quadrupoles; for simplicity, I
refer to it here as the polarization quadrupole.  It can also be
written as an IE,
\begin{equation}
     \Pi(\tau) = \Delta^{\rm T}_{2}(\tau) + 9 E_{2}(\tau),
\label{eqn:Pieqn}     
\end{equation}
with
\begin{equation}
     E_l(\tau) =
     \int_{\tau_i}^\tau \,
     d\tau'\, g(\tau,\tau')
     \frac{j_l( k(\tau-\tau'))}{ \left(k
     (\tau-\tau')\right)^2} \Pi(\tau').
\end{equation}
The CMB E-mode transfer function is then $\Delta^{\rm E}_l(\tau)
=(3/4)\sqrt{(l+2)!/(l-2)!} E_l(\tau)$.

\begin{figure*}
\centering
\subfigure[]{\includegraphics[width=.45\linewidth]{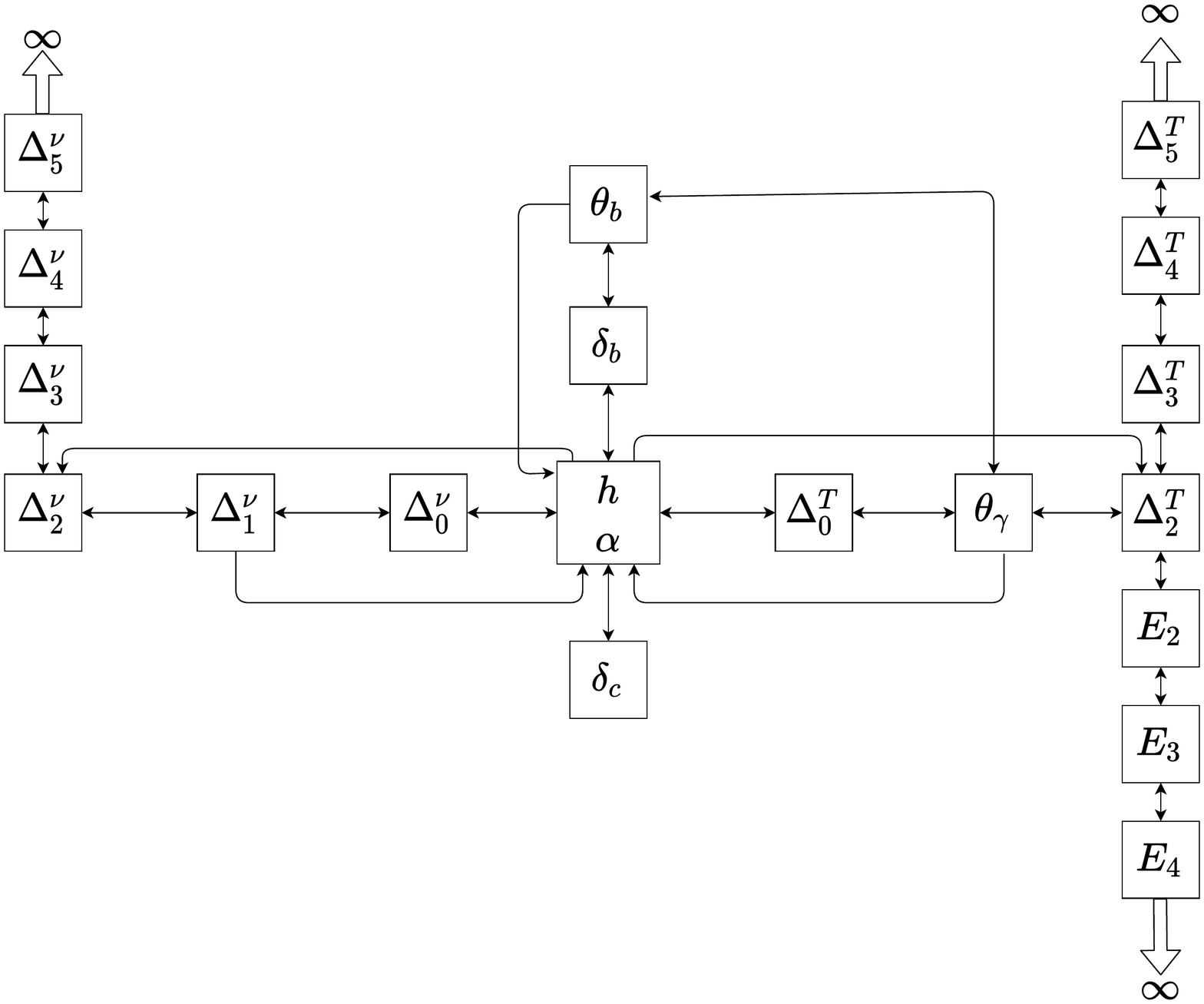}}
\quad
\subfigure[]{\includegraphics[width=.45\linewidth]{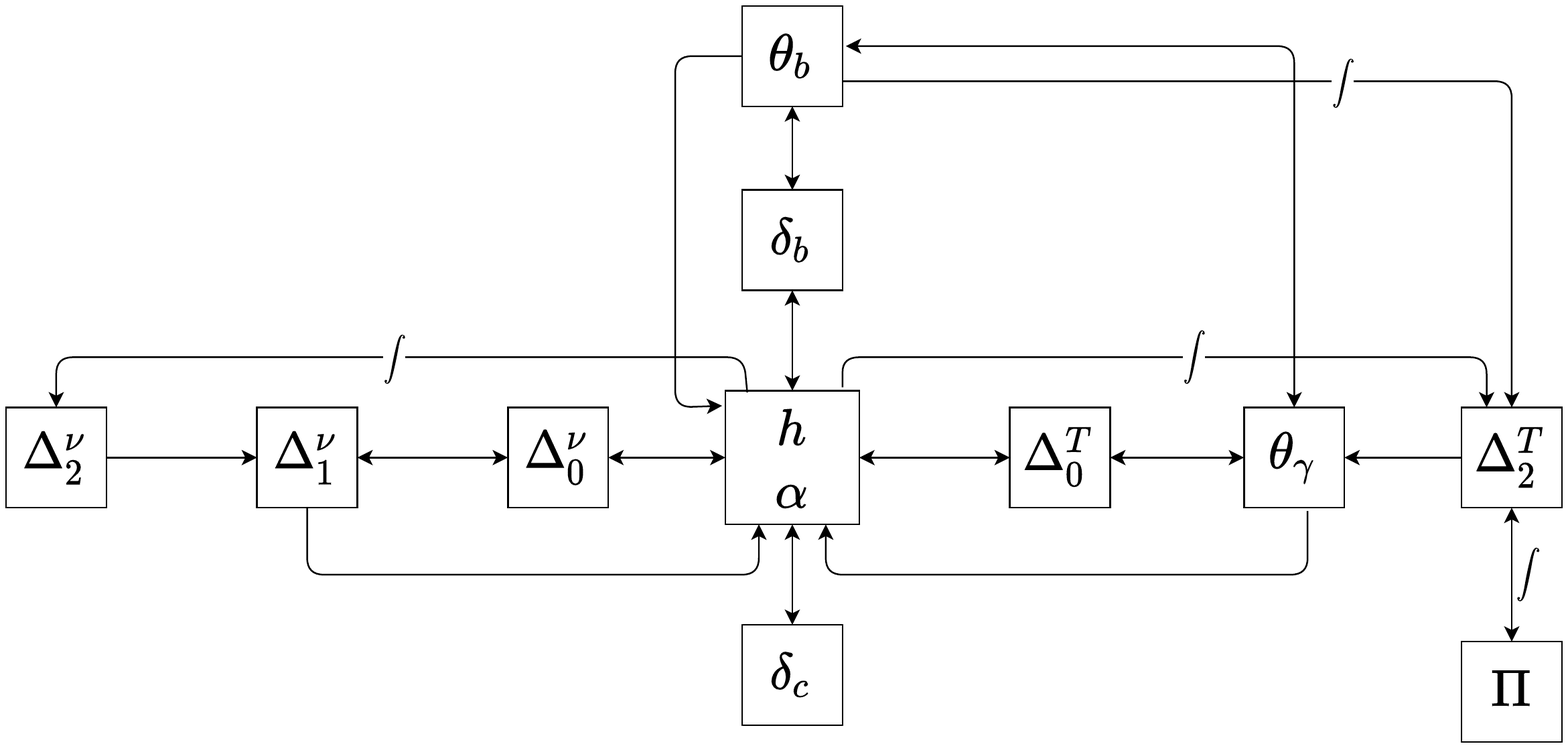}}
\caption{Flow charts for the perturbation calculation with (a)
     the Boltzmann hierarchy and (b) the integral-equation
     approach.  An arrow points from an element that appears
     in the differential equation for the element it points
     to.  Ingredients that appear in the integral equation for a
     given quantity are indicated in (b) with an integral sign.
     As both figures indicate, the higher moments ($l \geq3$ for
     $\Delta_l^{\rm T}$ and $\Delta_l^\nu$ and $\geq2$ for $E_l$)
     communicate to the rest of the system of equations only
     through the quadrupole ($l=2$).  The diagrams also indicate
     that in both cases, the photon-intensity quadrupole
     $\Delta_2^{\rm T}$ feeds into the rest of the system of
     equations only through the photon velocity $\theta_\gamma$,
     and similarly for the neutrino quadrupole.}
\label{fig:flowchart}
\end{figure*}

A derivation of Eqs.~(\ref{eqn:Tfluctuation}) and
(\ref{eqn:Pieqn}) will be provided in Ref.~\cite{inprep} using
the total-angular-momentum formalism \cite{Dai:2012bc}, but it
is easily verified that they agree with Eq.~(18) in
Ref.~\cite{Zaldarriaga:1996xe}, Eqs.~(74) and (77) in
Ref.~\cite{Hu:1997hp}, and with the IEs
in Ref.~\cite{Weinberg:2003ur}.  It can also be
verified, using the relation, $(2l+1)j_l'(x) = l j_{l-1}(x) -
(l+1) j_{l+1}(x)$ (which $R^{\rm LL}_l(x)$ and $j_l'(x)$ also
satisfy), that differentiation of these
two IEs recovers the usual Boltzmann hierarchy as
given, for example, in Eqs.~(2.4) of Ref.~\cite{Blas:2011rf} or
Eq.~(63) of Ref.~\cite{Ma:1995ey}.  Thus, these two IEs are
formally equivalent to the Boltzmann hierarchy.  For
completeness, the Boltzmann hierarchy is provided in the
notation/conventions used here in Appendix \ref{sec:boltz}.

\section{Implementation}
\label{sec:implementation}

The left
flowchart in Fig.~\ref{fig:flowchart} shows the interdependency
between the different perturbation variables in the differential
equations for their evolution.  In the middle are the
metric-perturbation variables $h$ and $\alpha$.  These are
sourced by the baryon, dark-matter, neutrino, and photon
densities and bulk velocities.  Apart from the baryon-photon
coupling that connects $\theta_\gamma$ and $\theta_b$, the only
communication between the different matter components is through
the metric perturbations.  The neutrino velocity is connected to
the neutrino quadrupole $\Delta^\nu_2$ which is then connected
to an infinite tower of Boltzmann equations for the higher-order
neutrino moments $\Delta^\nu_l$ for $l\geq3$.  The same can be
said for the photon velocity, except that there are two infinite
Boltzmann hierarchies for the higher photon-intensity and
photon-polarization moments.  When considered in tandem, the
photon monopole and dipole equations combine into a second-order
differential equation that resembles that for a driven simple
harmonic oscillater (discussed below); this describes oscillations of the
amplitude of the photon-baryon fluid driven by changes in the
metric perturbations and in the photon quadrupole.

In the line-of-sight approach \cite{Seljak:1996is}, the
Boltzmann hierarchy is solved up to a maximum multipole $l_{\rm
max}\sim 30$ to obtain the photon monopole, dipole, and
quadrupole, and $\Pi$ to reasonable accuracy.  The $C_l$ are
then obtained by evaluating the integrals in
Eqs.~(\ref{eqn:Tfluctuation}) and (\ref{eqn:Pieqn}).

As Fig.~\ref{fig:flowchart} illustrates, the 
two (nominally) infinite towers of photon differential
equations---one for the temperature moments ($\Delta_l^{\rm T}$
for $l\geq3$) and polarization moments ($E_l$ for
$l\geq2$)---communicate with the rest of the system of equations
only through the photon-intensity quadrupole $\Delta_2^{\rm
T}$.  Thus, one can replace the two photon Boltzmann hierarchies
with a pair of integral equations, one for $\Delta_2^{\rm T}$
and another for $\Pi$.  The rest of the system of equations is
then exactly the same as in the Boltzmann approach.

In this approach we retain the two lowest-order equations, for
the photon monopole ($l=0$) and dipole ($l=1$).  These equations are,
\begin{equation}
     \dot\Delta^{\rm T}_0 = -\frac13 \theta_\gamma - \frac16\dot
     h, \quad
     \dot \theta_\gamma = k^2 (\Delta^{\rm T}_0-2 \Delta^{\rm T}_2) - \dot\kappa
     \Theta_{\gamma b}, \label{eqn:thetagamma}
\end{equation}
with $\Theta_{\gamma b}(\tau) \equiv
\theta_\gamma(\tau)-\theta_b(\tau)$.  These equations are
supplemented by those,
\begin{equation}
     \dot\delta_b = -\theta_b -\frac12 \dot h, \quad
     \dot \theta_b = -{\cal H} \theta_b + c_s^2 k^2 \delta_b +
     \frac{\dot\kappa}{R} \Theta_{\gamma b},
\end{equation}
for the baryon density and velocity, respectively.  There is
also an equation, $\dot\delta_c = -\frac12 \dot h$,
for the CDM-density perturbation (the CDM peculiar velocity
vanishes in synchronous gauge).

The photon quadrupole $\Delta^{\rm T}_2(\tau)$ in
Eq.~(\ref{eqn:thetagamma}) is obtained at early times by the
TCA (up to second order in
$\dot\kappa^{-1}$, as described in
Refs.~\cite{CyrRacine:2010bk,Blas:2011rf} for improved
speed/precision).  The two equations for the early-time
evolution of $\theta_\gamma$ and $\theta_b$ can also be
replaced by their TCA, with the slip
$\dot\Theta_{\gamma b}$ evaluated (again, up to second order
$\dot\kappa^{-1}$) \cite{CyrRacine:2010bk,Blas:2011rf}.

At later times, the quadrupole is obtained from
Eq.~(\ref{eqn:Tfluctuation}) with $l=2$, along with
Eq.~(\ref{eqn:Pieqn}) for the time evolution of $\Pi(\tau)$.
With this approach, the equations in Eq.~(\ref{eqn:thetagamma})
combine to describe a driven oscillator damped by the photon
quadrupole \cite{semianalytic}.  The photon quadrupole is
provided at early times by the TCA and at later times from the
integral equation.

For completeness, the Einstein equations are 
\begin{equation}
     \ddot h +  \frac{\dot a}{a} \dot h = - 8 \pi G a^2 \left[ \delta \rho_{\rm tot} + 3 \delta p_{\rm tot} \right],
\label{eqn:firsteinstein}     
\end{equation}
\begin{equation}
    \frac13 (\dot h - \dot \alpha) = 8 \pi G a^2\left[ \frac43 \bar\rho_\gamma \theta_\gamma + \frac43 \bar \rho_\nu \theta_\nu + \bar \rho_{\rm b} \theta_{\rm b} \right],
\label{eqn:secondeinstein}    
\end{equation}
Note that the the Einstein equations are written here in terms
of the energy and momentum densities, but not the
anisotropic stress.  In this way, the photon-intensity
quadrupole $\Delta_2^{\rm T}(\tau)$ communicates with the rest of the
set of perturbation equations only through
Eq.~(\ref{eqn:thetagamma}).
The IEs for massless neutrinos are obtained from those
for photons, but setting $\Pi=\dot\kappa=0$.  These IEs have
come into play in the development of an effective
ultra-relativistic-fluid approximation \cite{Blas:2011rf}.

\section{Numerical solution of the integral equations}
\label{sec:numerics}

The IEs here are Volterra
equations of the second kind, which are typically solved as follows
\cite{Volterrabook,Press:1992zz}.   A pair of such equations has
the form,
\begin{equation}
     f^\alpha(t) = \int_a^t K^{\alpha\beta}(t,s) f^\beta(s) ds + g^\alpha(t).
\end{equation}
with $\alpha,\beta=1,2$ (and implied sum over repeated $\alpha,\beta$
not not $ij$).
They are solved on a mesh of $N$ uniformly spaced time steps
$t_i=a+ih$ with $i=1,2,\ldots,N$, with $h = (t-a)/N$.  The
integrals are then evaluated with the trapezoidal rule.
The solution to the IEs are then $f_{\alpha,0}=g_{\alpha,0}$ and
\begin{equation}
     \left(\delta^{\alpha\beta} - \frac12 h K^{\alpha\beta}_{ii}
     \right) f^\beta_i = h \left( \frac12
     K^{\alpha\beta}_{i0} f^\beta_0
     + \sum_{j=1}^{i-1} K^{\alpha\beta}_{ij} f^\beta_j\right) + g^\beta_i.
\label{eqn:volterrasolver}     
\end{equation}
For the pair of Volterra equations we deal with here,
the $2\times2$ matrix on the left-hand side must be inverted at
each time step \cite{Press:1992zz}.  The ordinary differential
equations, which must be solved simultaneously, are simply
stepped forward in time (i.e., Euler integration).

This algorithm works well if the kernels $K^{\alpha\beta}(t,s)$ are smooth and
slowly varying.  The visibility function in our integrands are
smoothly varying after decoupling begins to occur, at redshifts
$z\lesssim 1400$ ($\tau
\gtrsim 230$ Mpc).  The
perturbation variables that multiply it, as well as the radial
eigenfunctions, are also relatively smooth.  The trapezoidal-rule
integration therefore works reasonably well.  However,
for early conformal times ($\tau\lesssim 230$ Mpc) during tight
coupling, when $\dot\kappa \gg {\cal H}$, the visibility
function is very sharply peaked at
$\tau' \to \tau$.  The trapezoidal rule will therefore be
inaccurate (unless we take a huge number of time steps).

To remedy this, and to improve the transition
from tight coupling, we replace the trapezoidal rule in
$\Delta\tau'$ with one in $de^{-\kappa(\tau,\tau')}$.  More
precisely, we write the integrand in terms of the
visibility function, $(d/d\tau') e^{-\kappa(\tau,\tau')}$, times
the more slowly-varying perturbation variables.  The integrals
can then be written,
\begin{eqnarray}
   I(\tau) &=& \int^\tau \, d\tau' f(\tau') \frac{d}{d\tau'}\left[
   e^{-\kappa(\tau,\tau')} \right] f(\tau) \nonumber \\
   &\simeq& \sum_{n=1}
   \int_{\kappa_n}^{\kappa_{n-1}} d \left(
   e^{-\kappa(\tau,\tau')} \right)\left[ f_{n-1}
   \right. \nonumber \\ & & \left. \ \ \ +
   \left(\frac{df}{d\kappa'} \right)_{n-1}(\kappa-\kappa')\right],
\label{eqn:trapezoiddiff}   
\end{eqnarray}
where $\kappa_n=\kappa(\tau-nh)$, and $h$ is the small
conformal-time step.  The remaining $\kappa'$
integrals can then be done analytically and the derivative
$df/d\kappa'$ approximated by differencing.  Details are
provided in Appendix \ref{sec:details}.

By expanding the integrand $f(\tau)$ to linear order, as in
Eq.~(\ref{eqn:trapezoiddiff}), we obtain a result that
is exact for variations of $f(\tau)$ that are up to linear in
$\kappa$.  At early times, this then reproduces
the first-order TCA (to order
$\dot\kappa^{-1}$), even for one step that is not necessarily
small compared with $\dot\kappa^{-1}$.  The second-order TCA is
then recovered by evaluating the IE with two time steps.  This
allows a smooth transition from the TCA approximation to the IE
algorithm in Appendix \ref{sec:details}  as long as the TCA
values for the perturbation variables are stored for at least
two time steps.  At late times, the visibility function in
Eq.~(\ref{eqn:trapezoiddiff}) can be Taylor expanded to linear
order in $\Delta\kappa$.  Doing so then recovers the trapezoidal
scheme in Eq.~(\ref{eqn:volterrasolver}). 

The formula in Eq.~(\ref{eqn:volterrasolver}) requires for each
time step $i$ a sum over all earlier timesteps $j<i$.  However,
given the $e^{-\kappa(\tau,\tau')}$ factor in the visibility
function in the integrand, the sum can for all practical
purposes be started, for any given $\tau_i$ at some $j$ such
that $\kappa(\tau_i,\tau_j) \leq \Delta\tau_{\rm max} \simeq
10-20$.  If the other factors in the integrand are slowly
varying, this yields a precision degradation of $\lesssim
e^{-\Delta\tau_{\rm max}}$.

When the IE solver first begins, the
photon-baryon fluid is still tightly coupled, and so the
visibility function has support only over values of $\tau'$
fairly close to $\tau$; i.e., $(\tau-\tau') \lesssim N
\dot\kappa^{-1}$.  The argument $x=k(\tau-\tau')$ of the radial
eigenfunctions in Eq.~(\ref{eqn:Tfluctuation}) is thus small, and so the
radial eigenfunctions can be approximated as $j_2(x) \simeq x^2/15$,
$R^{\rm LL}_2(x) \simeq -1/5$, $j_2'(x) \simeq (2/15)x$.  The
integrand cannot, however, be approximated simply by the $R^{\rm
LL}(x)$ term, because $\Pi$ is ${\cal O}(\dot\kappa^{-1})$ times
$\theta_b$.  The third (i.e., the $\theta_b$) term contributes,
at lowest order in the TCA.

\section{Iterative solution of integral and differential
equations}
\label{sec:iterative}

The next step is to consider how to solve simultaneously the
differential equations for the rest of the system.  This
includes those for the metric-perturbation variables, and the
baryon and dark-matter densities and velocities.  It also in
principle includes the neutrino perturbation variables; here,
however, I will assume that these can be obtained with a
generalized-dark-matter \cite{Hu:1998kj} or
ultrarelativistic-fluid approximation (UFA) \cite{Blas:2011rf},
both of which have been made fairly effective.  In principle,
the integral-equation techniques described for photons here can
be applied to the neutrino sector as well.  For clarity, I focus
here, though, on the photon sector.

In trying to do so, however, the coupling between the IEs and
the DEs pose a chicken-and-egg problem:  The differential
equations for the rest of the system require knowledge of
$\Delta_2^{\rm T}(\tau)$, but the IEs for $\Delta_2^{\rm T}(\tau)$ cannot be
obtained without the solution to the DEs.  One
possibility is to solve the IEs and DEs simultaneously by
simply stepping the differential equations forward---i.e., Euler
integration.  This, however, requires very fine time steps,
especially toward the end of the TCA, and thus eliminates the
advantages of the early-time IE algorithm described above.
Another possibility is to step the IEs forward on a coarse time
grid, and then integrate the DEs forward (using an extrapolation
of the IE solutions from earlier time steps) using an
off-the-shelf adaptive-time-step DE solver.

However, the IEs and DEs can be solved very efficiently
with a simple iterative algorithm.  Here, we start with some
initial {\it anzatz} for $\Delta_2^{\rm T}(\tau)$ and $\Pi(\tau)$ and
then solve the DEs for all the other perturbation variables with
this {\it ansatz}.  We then integrate the IEs using the
solutions to those DEs to obtain new values of $\Delta_2^{\rm T}(\tau)$
and $\Pi(\tau)$.  We then iterate.  Of course, there is no
guarantee {\it a priori} that this iterative procedure will
converge to the correct answer, but some simple numerical
experiments show that this procedure converges, and does so
fairly quickly, even for a lousy (e.g., $\Delta_2^{\rm T}(\tau)=\Pi=0$)
initial {\it ansatz} for the IE solutions.

\begin{figure}[htbp]
\begin{centering}
\includegraphics[width=\columnwidth]{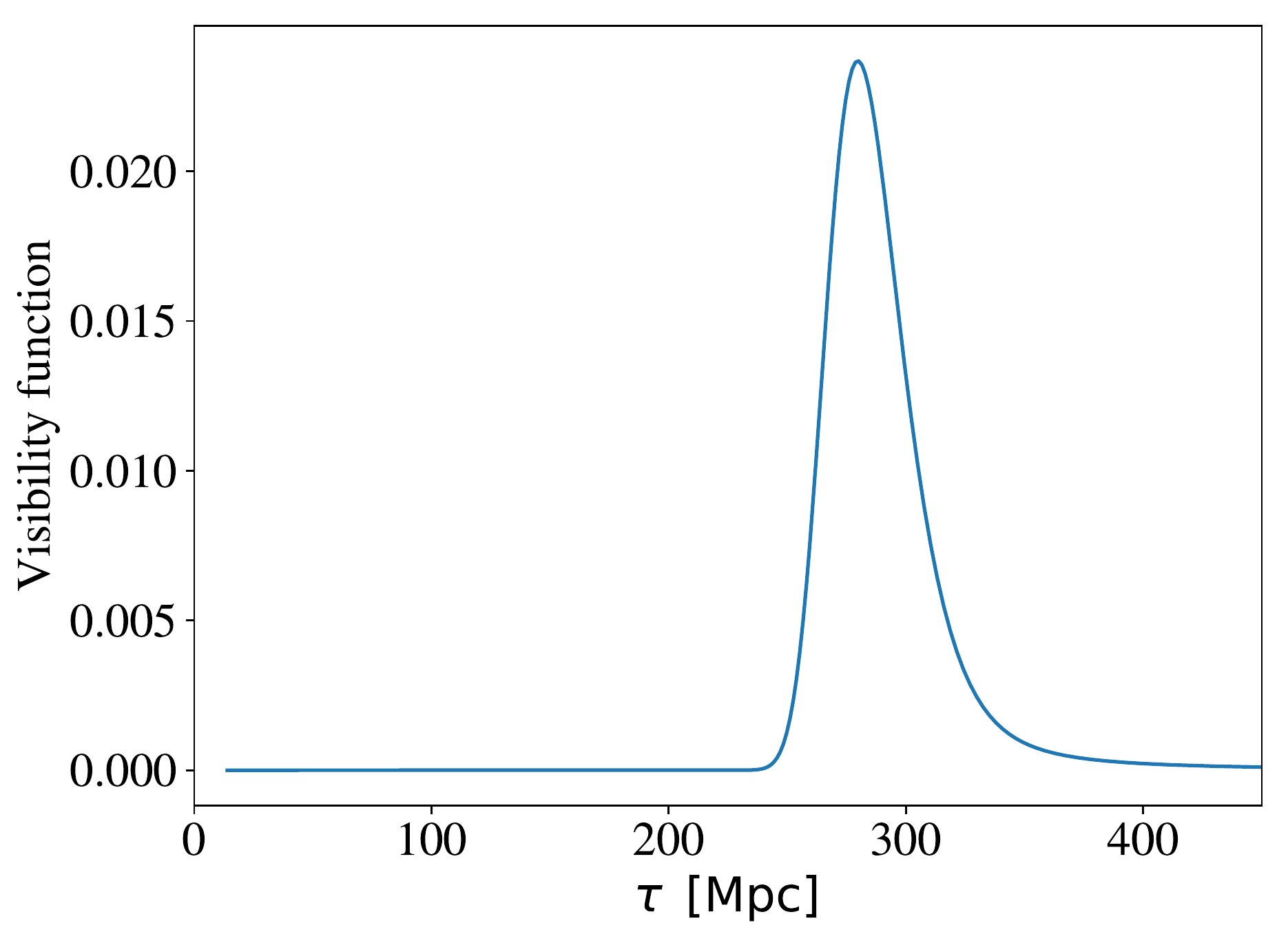}
\caption{The CMB visibility function $\dot\kappa(\tau_0,\tau)$
    as a function of conformal time.  It is shown to indicate
    the range of conformal times, peaked at
    $\tau\simeq280$ Mpc, that contribute to the observed
    CMB power spectra from recombination.}
\label{fig:vis}
\end{centering}
\end{figure}

\begin{figure}[htbp]
\begin{centering}
\includegraphics[width=\columnwidth]{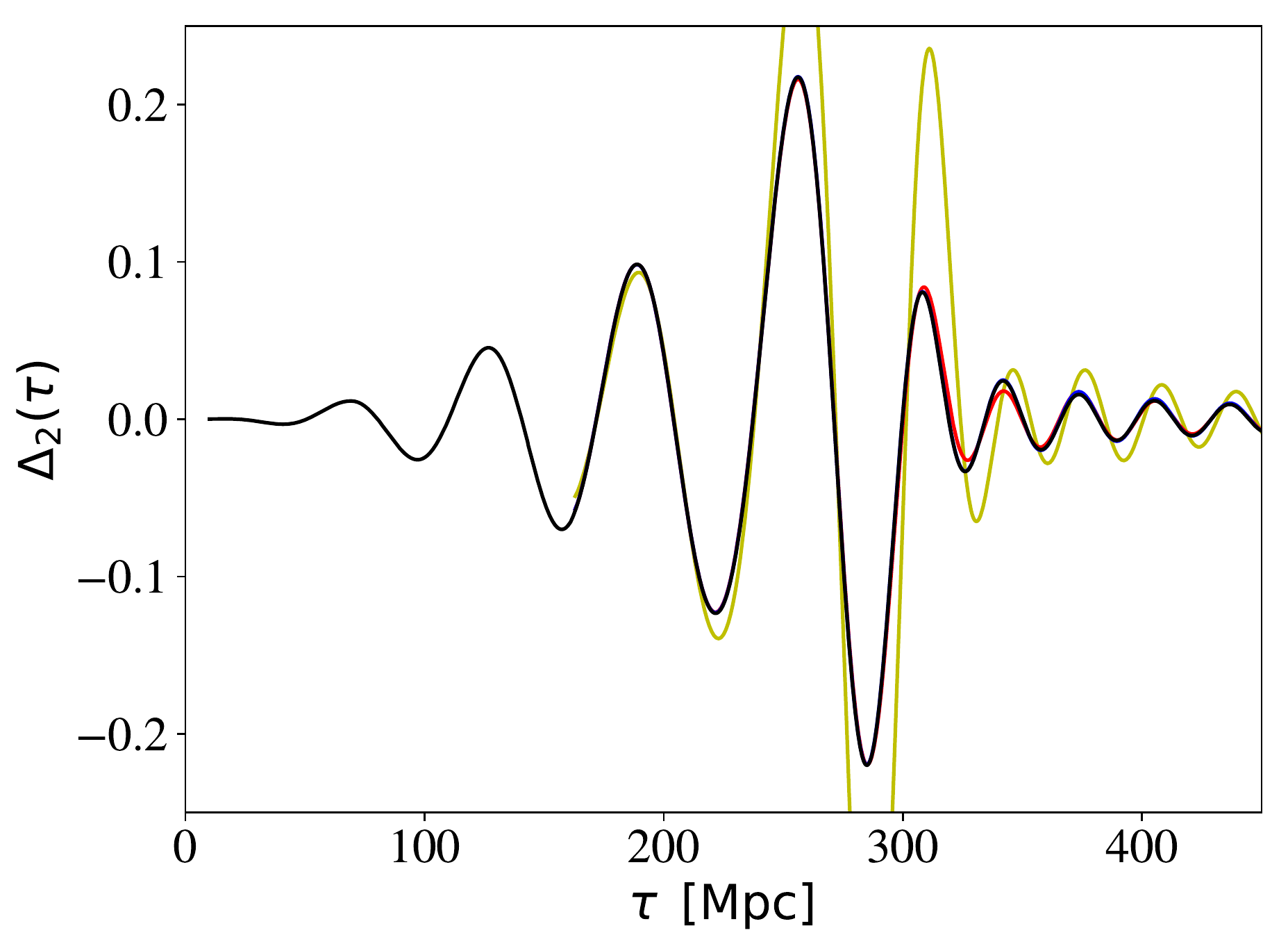}
\caption{The transfer function $\Delta_{2}^{\rm T}(\tau)$ for the CMB
    photon-intensity quadrupole as a function of conformal time
    $\tau$ for a Fourier mode of wavenumber $k=0.2$ Mpc
    (which corresponds roughly to a CMB multipole moment
    $l\sim3000$).  The black curve shows the results of the full
    Boltzmann hierarchy as a function of conformal time.  The
    other curves show results of the iterative integral-equation
    solution, taking $\Delta_2^{\rm T}(\tau)=0=\Pi(\tau)$ as an initial
    {\it ansatz}.  The yellowish curve shows the result for
    $\Delta_2^{\rm T}(\tau)$ after the first iteration---i.e., after
    integrating the differential equations for all perturbation
    variables except $\Delta_2^{\rm T}(\tau)$ and $\Pi(\tau)$ and then
    integrating the integral equations for $\Delta_2^{\rm T}(\tau)$ and
    $\Pi(\tau)$ using the results of the differential
    equations.  The red curve shows results after three
    iterations, and the blue curve after five iterations.  The
    thickness of the curves is such that if two are
    indistinguishable, the agreement between the two is
    $O(0.1\%$).}
\label{fig:Delta2}
\end{centering}
\end{figure}

\section{Numerical results}
\label{sec:results}

I have written a rudimentary C code to calculate the transfer
functions for the perturbation variables with the iterative
numerical implementation described here.
To simplify, I approximate neutrinos (taken to be massless) as a
generalized-dark-matter component with $w=c_s^2=c_{\rm
vis}^2=1/3$ \cite{Hu:1998kj}. I stop the code at redshift $z
\simeq 560$, after recombination but before reionization, and 
use an analytic approximation (which takes into account only
radiation and nonrelativistic matter at these times) for the
expansion history.  I use an ionization history from HyRec-2
\cite{hyrec}.  To compare this IE approach with the standard
Boltzmann hierarchy, I also wrote a second code that is
identical in every way except that it swaps out the integral
equations for $\Delta_2^{\rm T}(\tau)$ and $\Pi(\tau)$ for the
complete photon Boltzmann hierarchy.  The code uses an
off-the-shelf differential-equation solver \cite{Shampine} with
adaptive step size, although not necessarily optimized for stiff
equations.

In the IE code, the handoff from the TCA to the IE solver takes
place at $\tau=160$ Mpc.  The Boltzmann code uses the same TCA
at early times and then starts the full Boltzmann hierarchy at
$\tau=160$~Mpc.  The Boltzmann code
follows the Boltzmann hierarchy up to $l_{\rm max}=50$ (which I
found was required to keep the perturbation variables stable
over the $\tau$ range considered here).  The
results are similar, and the code a bit quicker, for smaller
$l_{\rm max}$.  The differential-equation solver in the
Boltzmann code runs with a relative error requirement of
$10^{-5}$ and absolute error of $10^{-4}$.  The integral
equations are evolved on a time grid that has spacing
$\Delta\tau=1.0$ from $160\, {\rm Mpc} \leq \tau \leq 240\, {\rm
Mpc}$ and $350\, {\rm Mpc} \leq \tau \leq 450\, {\rm
Mpc}$, and $\Delta\tau=0.5$~Mpc for $240\, {\rm Mpc} \leq \tau
\leq 350\, {\rm Mpc}$, for a total of 401 grid points.  The time
required for the IE part of the calculation scales as the square
of the number of grid points.

Fig.~\ref{fig:vis} shows the visibility function, which
indicates the conformal-time regime, $250\,{\rm Mpc} \lesssim
\tau \lesssim 400\,{\rm Mpc}$, over which the source functions
for the CMB power spectra are evaluated.

Fig.~\ref{fig:Delta2} illustrates the results of the numerical
experiment.  Shown there are results for the photon-intensity
quadrupole $\Delta_2^{\rm T}(\tau)$ of the Boltzmann code and
the iterative integral-equation results, starting from a naive
initial {\it ansatz} $\Delta_2^{\rm T}(\tau)=\Pi(\tau)=0$.  Results are
shown for $k=0.2$ Mpc, which corresponds roughly to CMB
multipole moments $l\sim3000$, near the upper limit of current
measurements.  The frequency of oscillations in the transfer
function are reduced at smaller $k$, and so the numerical
algorithm should, if anything, work even better at lower $k$.

The results are shown for one iteration (yellow), three iterations (red) and
(five iterations) blue.  The iterative solutions converge first
at early times and then require more iterations to converge at
later times.  The overlap between the black and blue (5
iterations) curves indicates that the agreement is at the
$O(0.1\%)$ level over the conformal-time range that contributes
to the observed CMB power spectra.  This IE code takes
$\sim0.15$ times as long to run as the Boltzmann code, implying
that each iteration can be completed in $\sim1/30$ the time
required for the Boltzmann code.  Both codes are fairly
rudimentary, and so these time comparisons should be taken with
a grain of salt.  Still, these results suggest that this may
provide a route to speeding up the standard Boltzmann codes.

There may be room for even further improvement.  The results
shown in Fig.~\ref{fig:Delta2} are obtained using the most naive
possible initial {\it ansatz} for $\Delta_2^{\rm T}(\tau)$ and
$\Pi(\tau)$.  The number of iterations required for convergence
to the required precision can be reduced if one starts with a
better initial guess for these quantities.  It should be
possible to derive a simple semi-analytic {\it ansatz} that
interpolates between the well-understood early-time TCA behavior
and the late-time behavior, which comes from the Sachs-Wolfe
effect.

One should, however, be able to do even better.  These
calculations are not performed in isolation.  In cosmological
MCMC analyses, the Boltzmann codes are run repeatedly to map the
likelihood functions in a multidimensional
cosmological-parameter space.  Thus, each time the calculation
is done, it has presumably already been done for a nearby point
in that cosmological parameter space.  Thus, it should be
possible to start the iterative algorithm by using the results
for $\Delta_2^{\rm T}(\tau)$ and $\Pi(\tau)$ from a previous run.  To
test this, I ran the code using as the initial {\it ansatz} the
results for $\Delta_2^{\rm T}(\tau)$ and $\Pi(\tau)$ from a prior run
with $\Omega_b$ reduced by 2\%.  This code converges to
$O(0.1\%)$ after just one iteration.

\section{Conclusions and ideas for future work}
\label{sec:conc}

I have presented an alternative formulation of the equations
for the evolution of cosmological perturbations in which the
infinite Boltzmann hierarchy for the photon distribution
function is replaced by a pair of integral equations.  There is
no new physics here---it is simply a recasting of the equations
in a way that may lead to physical insight and alternative
schemes for numerical solution.  As was known from the
line-of-sight approach \cite{Seljak:1996is}, CMB fluctuations
are determined only by the photon monopole (energy density),
dipole (peculiar velocity), and quadrupole (more specifically,
$\Pi$).  In the Boltzmann hierarchy, these are the result of
some complicated transfer of power between these lower moments
of the photon distribution function and an infinite tower of
higher moments.  The IE formalism shows, however, that the
lower moments, and in particular the quadrupole moment, at the
surface of last scatter (i.e., those that enter into the
line-of-sight integration) are simply described by the exact
same equations that describe the lower moments that we see.

I have shown that simple iterative solution of the combined
system of integral and differential equations does a pretty good
job at reproducing the results of the Boltzmann calculation in a
fraction of the time.  This exercise also shows that the IE
formalism can be implemented numerically without
(apparently) any significant numerical instabilities---this was
not a foregone conclusion, given the occurrence of instabilities
in some IE solvers \cite{Volterrabook}, as well as those that may
arise from finite $l_{\rm max}$ in the Boltzmann hierarchy.

There is, however, far more work that needs to be done before we
know whether this approach can implemented to speed up a code
like CLASS or CAMB.  These codes benefit from a number of
insights and clever algorithms, whereas what I have presented
here is fairly naive.  Those codes also have controlled errors,
whereas the grid spacing in my calculation was guessed to provide
an $O(0.1\%)$ precision in $\Delta_2^{\rm T}(\tau)$.

The spacing of the conformal-time grid points in the
integral-equation solver is an obvious thing to explore.  
In this calculation I simply estimated the number of grid points
that would be required for $O(0.1\%)$ precision.  However, the
distribution of grid points can certainly be optimized to
provide the desired observables (e.g., CMB and matter power
spectra) to the required precision.  Good results can probably
also be obtained for smaller $k$ with fewer grid
points, given the smoother integrands at lower $k$.  The current
code also sums over all prior grid points.  However, given the
high opacity at early times, the sum can be restricted only to
grid points that are at an optical depth $\Delta\kappa \lesssim
5$ earlier.  There are also algorithms, more sophisticated than
the trapezoidal-rule algorithm used here, on numerical
solution to Volterra equations (e.g., Ref.~\cite{Baker:2000}) in
the literature that may be worth exploring.  Finally, there may
be alternative implementations of the integral/differential
equations that may be better suited for numerics.  For example,
it should be possible to eliminate the differential equations
for the photon monopole and dipole and replace the integral
equation for the quadrupole $\Delta_2^{\rm T}(\tau)$ with that
for the monopole $\Delta_0^{\rm T}(\tau)$.  Or perhaps the
differential equation for $\Delta_2^{\rm T}(\tau)$ can be
included and the integral equation replaced by one for
$\Delta_3^{\rm T}(\tau)$.

\begin{acknowledgments}
I thank L.~Ji, R.~Caldwell, D.~Grin, J.~Bernal, and E.~Kovetz
for useful discussions and comments on an earlier draft.  This
work was supported by NSF Grant No.\ 1818899 and the Simons
Foundation.
\end{acknowledgments}

\appendix

\section{Boltzmann hierarchy}
\label{sec:boltz}

For completeness and comparison with prior work, I provide the
Boltzmann equations for the photon moments in the notation used
here.  These equations are derived by differentiating
Eqs.~(\ref{eqn:Tfluctuation}) and (\ref{eqn:Pieqn}) with
respect to $\tau$.  The independent variable $\tau$ appears in
the limit of integration, the visibility function, and in the
radial eigenfunctions, and all of the radial eigenfunctions
satisfy the spherical-Bessel-function relation,  $(2l+1)j_l'(x)
= l j_{l-1}(x) - (l+1) j_{l+1}(x)$.  The monopole and dipole
equations are already provided in Eq.~(\ref{eqn:thetagamma}).
The equations for $l\geq2$ are
\begin{eqnarray}
     \dot\Delta^{\rm T}_l &=& - \dot\kappa \Delta^{\rm T}_l +
     \frac{kl}{2l+1} \Delta^{\rm T}_{l-1} -
     \frac{k(l+1)}{2l+1} \Delta^{\rm T}_{l+1}  \nonumber\\
     & & \ \ \ \ \ \ +\frac15 \left(\frac{\dot\alpha}{3} + \frac{\dot\kappa
     \Pi}{2} \right) \delta_{l2}, \nonumber \\
     \dot E_l & = & -\dot\kappa E_l + \frac{k(l-2)}{2l+1} E_{l-1} -
     \frac{k(l+3)}{2l+1} E_{l+1} +\frac{1}{15} \dot\kappa \Pi
     \delta_{l2},\nonumber\\
\label{eqn:boltzmann}
\end{eqnarray}
with $\Pi=\Delta^{\rm T}_l + 9 E_l$.

\section{Details of the IE solver}
\label{sec:details}

We first define functions $I^{\rm T}(\tau,\tau')$ and
$I^\Pi(\tau,\tau')$ by writing
\begin{eqnarray}
     \Delta^{\rm T}_2(\tau) &=& \int^\tau d\tau' g(\tau,\tau')
     I^{\rm T}(\tau,\tau'). \nonumber \\
     \Pi(\tau) &=& \int^\tau d\tau' g(\tau,\tau')
     I^\Pi(\tau,\tau').
\end{eqnarray}
The integrals are then discretized, taking into account the fact
that $\Pi(\tau)$ appears in $I^{\rm T}(\tau,\tau')$ and
$I^\Pi(\tau,\tau')$, in the following way.  We define two
sums,
\begin{eqnarray}
     \Delta^0_{2,i+1} &=& \sum_{j\leq i} \left( I^{\rm T}_{j+1}
     W^+_j + I^{\rm T}_j
     W_j \right) - \frac{1}{10} \Pi_{j+1} W^+_i \nonumber\\
     \Pi^0_{i+1} &=& \sum_{j\leq i} \left( I^\Pi_{j+1} W^+_j + I^\Pi_j
     W_j \right) - \frac{3}{5} \Pi_{i+1} W^+_j,\nonumber
\end{eqnarray}
where $\Pi_i = \Pi(\tau_i)$, $I^{\rm T}_j=I^{\rm
T}(\tau_{i+1},\tau_j)$, and $I^\Pi_j=I^\Pi(\tau_{i+1},\tau_j)$.
Here the weight functions are 
\begin{eqnarray}
     W^+_j &=& e^{-\kappa(\tau_{i+1},\tau_{j+1})}\left(1 -
     e^{-\Delta\kappa_j} - \frac{1 -
     (1+\Delta\kappa_j)e^{-\Delta\kappa_j}}{\Delta\kappa_j} \right), \nonumber \\
     W_j & = & \frac{e^{-\kappa(\tau_{i+1},\tau_{j+1})}}{\Delta
     \kappa_j} \left[1 -
     (1+\Delta\kappa_j)e^{-\Delta\kappa_j}
     \right]
\end{eqnarray}
where $\Delta\kappa_j = \kappa(\tau_{j+1})-\kappa(\tau_j)$.
These weight functions approach $W^+_j \to \Delta\kappa_j/2$ and
$W_j \to \Delta\kappa_j/2$ at late times, thus recovering 
Eq.~(\ref{eqn:volterrasolver}) (written as an integral over
$\kappa$, rather than $\tau$).  At early times, $W_j^+ \to 1 -
(\Delta\kappa)^{-1}$ and $W^j \to (\Delta\kappa)^{-1}$; this
then recovers the first-order tight-coupling approximation,
$\Delta_2 =(2/5) \Pi= (4/45)(\dot\alpha+2\theta_b)
/\dot\kappa$, even from one time step in the evaluation of the
integral---the second-order TCA is reproduced by two time steps.

The discretized quadrupoles are then,
\begin{eqnarray}
     \Pi_{i+1} &=& \frac{\Pi^0_{i+1} + \Delta^0_{2,i+1}}{ 1-
     \frac{7}{10} W^+_i}, \nonumber \\
     \Delta^{\rm T}_{2,i+1} &=& \Delta^0_{2,i+1} + \frac{1}{10}
     \Pi^0_{i+1} W^+_i.
\end{eqnarray}

\end{document}